\documentclass[12pt]{article}
\usepackage{xcolor}
\usepackage{amssymb}
\usepackage{amsmath}
\usepackage{amsfonts}
\usepackage{graphicx}
\usepackage{mathrsfs}
\usepackage{comment}
\usepackage{cite}
\usepackage{authblk}
\usepackage{array}
\usepackage{subcaption}

\newcommand{\Slash}[1]{{\ooalign{\hfil#1\hfil\crcr\raise.167ex\hbox{/}}}}

\newcommand{\beq}{\begin{equation}}  \newcommand{\eeq}{\end{equation}}
\newcommand{\bef}{\begin{figure}}  \newcommand{\eef}{\end{figure}}
\newcommand{\bec}{\begin{center}}  \newcommand{\eec}{\end{center}}
  
\newcommand{\laq}[1]{\label{eq:#1}}  

\newcommand{\Eq}[1]{Eq.(\ref{eq:#1})}

\newcommand{\ab}[1]{\left|{#1}\right|}
\newcommand{\vev}[1]{ \left\langle {#1} \right\rangle }

\def\({\left(}
\def\){\right)}

\def\O{\mathcal{O}}

\newcommand{\AND}{~{\rm and}~}
\newcommand{\EV}{ {\rm \, eV} }
\newcommand{\KEV}{ {\rm \, keV} }
\newcommand{\MEV}{ {\rm \, MeV} }
\newcommand{\GEV}{ {\rm \, GeV} }

\def\f{\phi}
\def\g{\gamma}

\def\D{\Delta}

\def\L{\Lambda}

\def\*{\dagger}

\usepackage[normalem]{ulem}

\begin{document}
\begin{titlepage}
\begin{center}

\vspace{1.0cm}
{\Large\bf Direct Detection of Cosmic Walls\\ with Paleo Detectors
}
\vspace{1.0cm}

{\bf  Wen Yin}

\vspace{1.0cm}
{\em 
$^{1}${Department of Physics, Tokyo Metropolitan University, Tokyo 192-0397, Japan\\}
}

\vspace{1.0cm}
\abstract{ 
Paleo detectors are emerging dark matter  detection technology that exploits ancient minerals as passive, time-integrated detectors.  
Unlike conventional real-time experiments, they search for permanent damage tracks—typically tens of nanometers to micrometers long—left in crystal lattices by rare particle interactions, most notably dark matter induced nuclear recoils accumulated over millions to billions of years.  In this paper I propose a direct detection strategy for cosmic walls—either bubble walls produced by a late-time first-order phase transition or domain walls in a scaling regime—using paleo detectors as the target medium.  
Because the cosmic wall is expected to traverse Earth at most $\mathcal{O}(1)$ time(s) in cosmic history, an ancient, continuously exposed detector is the only feasible way to observe it directly.  By calculating the target recoils, I find that the  
 smoking-gun signature would be parallel damage tracks found worldwide in minerals older than the wall-crossing epoch. I derive the  limit on the wall-target coupling by assuming that a wall passed through the Earth within the last 0.5 Gyr.
 I also mention a novel indirect detection of ultra-relativistic cosmic walls by noting the induced cosmic rays. 
}

\end{center}
\end{titlepage}

\setcounter{footnote}{0}
\section{Introduction}

Recent years have seen growing interest in the use of paleo detectors%
\cite{Fleischer1964,Fleischer,Fleischer:1965yv,Snowden-Ifft:1995zgn,Baum:2023cct}—ancient natural minerals that serve as passive, long-duration solid state nuclear track detectors—offering a novel approach complementary to conventional real-time experiments for dark matter or neutrinos.
The central idea is that nuclear recoils induced by them leave damage tracks in the crystal lattice of insulating minerals, which can survive over geological timescales of hundreds of millions to billions of years.
These tracks, typically nanometer- to micrometer-scale in length, can now be imaged with techniques such as small angle X-ray scattering, helium ion microscopy, or atomic force microscopy.

Because of their extremely long integration times and the directional information encoded in track morphology, paleo detectors are sensitive to rare interactions involving sub-GeV to multi-TeV dark matter and other non-standard scenarios.
Their natural shielding from cosmic rays and very low intrinsic radioactivity also make them ideal for ultra-low-background searches.
Feasibility studies with minerals such as muscovite, olivine, halite, and apatite have been reported in Refs.\,\cite{Edwards:2018hcf,Baum:2018tfw,Drukier:2018pdy}.
Although challenges remain—chiefly high-resolution, high-throughput read-out, track-energy calibration, and radiogenic-background modelling—ongoing research and development indicates that paleo detectors could substantially extend our reach for elusive dark matter signals and other cosmological relics.

Bubble walls and domain walls, on the other hand, may survive into the late-time Universe.
A first-order phase transition (FOPT) occurs in extensions of the Standard Model: bubbles nucleate by quantum tunnelling, their walls expand, and eventually collide to complete the transition.
Domain walls arise as topological defects in  scenarios with degenerate vacua; once produced, the walls rapidly enter a scaling regime with $\O(1)$ wall per Hubble horizon at each epoch.
I refer to both kinds of defects collectively as {\bf cosmic walls}.

Several theoretical ideas motivate such walls in the late-time Universe.
For example, modest variations in the electron mass or the fine-structure constant, possibly generated by a late FOPT, can improve fits to cosmological data and alleviate existing tensions~\cite{Sekiguchi:2020teg,Toda:2025dzd}.
Axion-like-particle domain walls provide a natural explanation for the reported $4\sigma$ cosmic birefringence signal~\cite{Takahashi:2020tqv,Lee:2025yvn}.

Because a wall is expected to cross Earth at most $\mathcal{O}(1)$ time over cosmic history, searches usually focus on indirect signatures, i.e. indirect detection in the dark matter terminology,—cosmic birefringence~\cite{Agrawal:2019lkr,Takahashi:2020tqv,Lee:2025yvn,Minami:2020odp,Louis:2025tst}, gravitational waves from wall collapse (see e.g. the recent nanoGrav explanation \cite{NANOGrav:2023gor,Kitajima:2023cek,Bai:2023cqj,Blasi:2023sej,Gouttenoire:2023ftk,Ferreira:2024eru,Gruber:2024jtc,Antoniadis:2023ott,Reardon:2023gzh,Xu:2023wog}), or gravitational effects on the cosmic microwave background~\cite{Zeldovich:1974uw,Vilenkin:1984ib,Ferreira:2023jbu}.
Direct detections are rarer  to be studied; the GNOME magnetometer network~\cite{Masia-Roig:2019hsy,GNOME:2023rpz,Pospelov:2012mt}, for instance, targets models with domain wall densities much larger than the scaling prediction~\cite{Battye:1999eq,Avelino:2006xy,Friedland:2002qs}. 

In this paper, I propose using paleo detectors to perform a direct detection search for cosmic walls.
I compute the recoils induced when a wall passes through a mineral in the weak-coupling limit and show that even very feebly interacting scenarios can already be constrained.
Since the wall crossing is at most a once-per-cosmic-history event, an ancient, continuously exposed detector provides the only realistic means of observing it directly.

\section{Recoils from cosmic walls}

Before discussing specific wall-formation scenarios, I first outline how a cosmic wall interacts with scatterers inside a paleo detector.  
For concreteness, assume that the wall field $\phi$ couples to a Dirac fermion $N$—which can be a nucleon,\footnote{The extension to the case that the nuclei is a scalar gives the same result. See the Appendix.  
} an atomic nucleus, or an electron—through  
\beq\laq{int}
  \mathcal{L}_{\mathrm{int}} = - f(\phi) \, \bar N N ,
\eeq
where $f$ is a generic function to be specified later.  
Such a coupling can be UV-completed, for example, if $\phi$ mixes with the Standard-Model Higgs via a portal interaction~\cite{Sakurai:2021ipp}.
 Assuming $\f$ couples to the fundamental nucleon universally with $f^{n}(\f) \bar n n $, I expect $f(\f)\simeq A f^{n}$ with $A$ being the mass number of the nuclei because of the mass scaling. 
This implies various calculations on nuclei can be reduced to the study on nucleon recoils as long as the low energy effective theory for the Lagrangian is valid. From the similar reason, I expect that when the nuclei is a boson whose mass scales with $A$, we can get a similar conclusion. This is numerically  shown in the Appendix.  
 
Expand the field around its vacuum value $\vev{\phi}=v$ by writing  
$\phi = v + s / \sqrt{2}$.  
The fluctuation $s$ acquires a Yukawa interaction $y = f'(v)$; existing constraints require $y \ll 1$ (see, e.g.~\cite{Yin:2024txg}). From this reason and to simplify the discussion, I consider the mass shift of $N$ is negligible and I ignore it throughout the paper.

I assume that the wall has a characteristic width $1/m_{\phi}$, where $m_{\phi}$ is the scalar mass.
A convenient smooth profile is
\beq
\phi(z) = v \frac{1 - \tanh(m_{\phi} z)}{2},
\eeq
but any smooth choice gives the same qualitative behavior.
I orient the $z$-axis along the wall’s motion through Earth and denote the wall velocity in the Earth frame by $v_w$, which I take to be comparable to or larger than the typical velocity of $N$.
Over the detector scale the wall is treated as perfectly planar, so its curvature is neglected.

For a wall to survive well after recombination I impose the condition that solves the ``domain wall problem’’\footnote{If the wall is highly Lorentz-boosted this bound becomes even stronger.}  
\beq
  v^2 m_{\phi} \lesssim \MEV^3 ,
\eeq
as in Refs.\,\cite{Zeldovich:1974uw,Vilenkin:1984ib}.  
This guarantees that the wall’s gravitational field does not spoil cosmological observations.

A perturbative bound follows from the self-coupling of the scalar:
\beq
  m_{\phi} \lesssim v .
\eeq
One needs a potential barrier in the interval $\(0,v\)$; if $m_{\phi}$ exceeded $v$, the curvature of the potential would vanish inside this range, forcing higher derivatives—and hence self-couplings—to exceed unity in units of $m_{\phi}$.  
Combining the two bounds gives\footnote{If this limit is relaxed in specific model constructions, the thin-wall regime, $2\pi/|p_z| \gg 1/m_{\phi}$ may be allowed.  Then the wall can reflect $N$ with probability $\simeq \frac14|\Delta m_N m_N/(p_z^{w})^2|^2$, giving reflected momentum $2p_z^{w}$ from the simple discussions on the boundary conditions. 
Here $\D m_N$ is the tiny mass difference of $N$ before and after the wall passage.
  For $p_z^{w}\sim100\MEV$ one needs $\Delta m_N\gtrsim7\times10^{-14}\GEV$ to obtain more than $\O(10)$ tracks for $N=10^{24}$ targets.}
\beq
  m_{\phi} \lesssim 1 \MEV .
\eeq

In the wall frame $N$ at rest in Earth moves with velocity $-v_w$, giving longitudinal momentum
\beq
  |p_z^{w}| \simeq \gamma_w v_w m_N ,
\eeq
where $\gamma_w = 1/\sqrt{1-v_w^2}$ is the wall Lorentz factor.

Because $|p_z^{w}|\gtrsim \MEV$ and the wall width is $1/m_{\phi} > 1 \MEV^{-1}$, the system is in the thick-wall limit $2\pi/|p_z| \ll 1/m_{\phi}$.  
In this limit the wall can be treated as a background field following a WKB approximation
 and reaction rates can be estimated from perturbation theory, following well-established techniques developed for expanding bubble walls in the early Universe, such as the study of wall friction~\cite{Bodeker:2009qy,Bodeker:2017cim,Azatov:2020ufh,GarciaGarcia:2022yqb}, dark matter productions,\cite{Azatov:2021ifm,Baldes:2022oev,Azatov:2022tii,Azatov:2023xem,Azatov:2024crd,Ai:2024ikj}, 
 and particle production for baryogenesis~\cite{Azatov:2021irb,Baldes:2021vyz,Azatov:2023xem,Chun:2023ezg}.
In particular, particle production via the higher dimension term from the interaction with the wall was studied in \cite{Azatov:2024crd}.

First, the elastic process $N\to N$ produces practically no recoils in $N$.  
Energy conservation together with conservation of the $x$- and $y$-components of momentum forces the $z$-momentum of the final state to coincide with that of the initial state; the tiny mass shift induced by \Eq{int} is negligible, so the net change in $p_z$ is vanishing.\footnote{%
  If one evade the scalar constraints, a large shift in $m_N$ could in principle reflect the low-momentum mode of $N$ in the wall frame. For instance, a domain wall whose interior gives $N$ a much larger mass allows passage via tunneling; the reflected component would then damage the detector for the region that the tunneling is dominant.  Such a scenario might be relevant to certain cosmic tensions and to models of cosmic birefringence, but its detailed study is left for future work. }

The leading inelastic channel is therefore
\beq
  N(p_i)\;\to\;N(p_f)\,s(p_{\phi}) ,
\eeq
where, in the wall frame,
\beq
  p_i = \{p_i^0,0,0,p_i^z\},\quad
  p_f = \{p_i^0(1-x), p_{\perp},0,p_f^z\},\quad
  p_{\phi} = \{p_i^0 x, -p_{\perp},0,p_{\phi}^z\}.
\eeq
Without loss of generality the transverse momentum $p_{\perp}$ is chosen along the $x$-axis.  
Define
\beq
  \Delta p_z \;=\; p_f^z + p_{\phi}^z - p_i^z
  = \sqrt{p_0^2(1-x)^2 - p_{\perp}^2 - m_N^2}
    + \sqrt{p_0^2 x^2 - p_{\perp}^2 - m_{\phi}^2}
    - \sqrt{p_0^2 - m_N^2}\!.
\eeq

The transition probability is~\cite{Azatov:2024crd}
\begin{align}
  P_{N\to N\phi}
  &= \frac12 \sum_{\text{spins}}
     \int \frac{d^3 p_f\,d^3 p_{\phi}}
              {(2\pi)^3\,2p_f^0\,2p_{\phi}^0\,2p_i^0}\;
     \delta^{(3)}(p_f + p_{\phi} - p_i)\;
     \bigl| F \bigr|^2\,
     \frac{|\mathcal{M}|^2}{|\Delta p_z|^2}  \\[4pt]
  &= \frac12 \sum_{\text{spins}}
     \int \frac{dx}{8 p_f^z p_{\phi}^z}
     \int \frac{dp_{\perp}^2}{8\pi^2}\;
     \frac{|\mathcal{M}|^2}{(\Delta p_z)^2}\,
     |F|^2 , \laq{int}
\end{align}
where the squared matrix element from the $s\,\bar N N$ vertex is
\beq
  \sum_{\text{spins}} |\mathcal{M}|^2
  = 4\!\left(p_f\!\cdot\!p_i + m_N^2\right)
  = 4\!\left[
      p_0^2(1 - x)
      - \sqrt{p_0^2 - m_N^2}\,
        \sqrt{p_0^2(1 - x)^2 - p_{\perp}^2 - m_N^2}
      + m_N^2
    \right].
\eeq
The filtering function
\beq
  F = \int dz\;e^{i \Delta p_z z}\,
      \partial_z f'\!\bigl( \vev{\phi(z)} \bigr)
\eeq
controls the allowed violation of $p_z$ conservation.  
In two useful limits irrelevant to the detail form of $F$ and domain wall profile,
\begin{align}\laq{F}
  F \;\longrightarrow\;
  \begin{cases}
    0 , & \text{if } \ab{\Delta p_z} \gg m_{\phi}, \\[4pt]
    f'(v) - f'(0) , & \text{if } \ab{\Delta p_z} \ll m_{\phi}.
  \end{cases}
\end{align}

We can also estimate averaged recoil observables by
\beq
  \vev{X}
  \;\approx\;
  P^{-1}_{N\to N\f}\,
  \left(
    \frac12\sum_{\text{spins}}
    \int\frac{dx}{8\,p_f^z p_\f^z}
    \int\frac{dp_\perp^{2}}{8\pi^{2}}\,
    \frac{|\mathcal{M}|^{2}}{(\Delta p_{z})^{2}}\,
    |F|^{2}\,X
  \right),
\eeq
with
$X = p_{f,z}^{\text{Earth}},
      \,(p_{f,z}^{\text{Earth}}-\vev{p_{f,z}^{\text{Earth}}})^{2},
      \,p_\perp^{2}$,
etc. Here the superscript of Earth denotes the value in the Earth frame.

\section{Results}

\subsection{Bubble walls from late-time FOPT}

The first class of cosmic walls of interest is  \emph{bubble walls}
Consider a FOPT that completes after the minerals acting as paleo detectors were formed.  
Bubbles nucleate in the false vacuum; their boundaries are energetic walls, which accelerate until they reach a terminal Lorentz factor $\g_w$ set by balance between the driving force from the difference of vacuum energy and friction due to ambient particles.  
Subsequent bubble collisions fill the Universe with the true vacuum. 

With probability $\O(1)$ a wall intersects a given mineral sample at speed $\g_w$.  
Because the wall size is macroscopic compared with the sample, curvature can be neglected and the formalism of the previous section applies.  
Momentum transfer to nucleons, nuclei or electrons then leaves charged tracks in the mineral.

For this scenario I adopt
\beq
  f(\f) \;=\; \frac{|\f|^{2}}{M},
\eeq
assuming that $\f$ carries a charge under a symmetry not shared by $N$.  
This yields a position-dependent effective Yukawa coupling $f'(\f)\propto\f/M$.
   \begin{figure}[t!]
  \begin{center}  
\includegraphics[width=140mm]{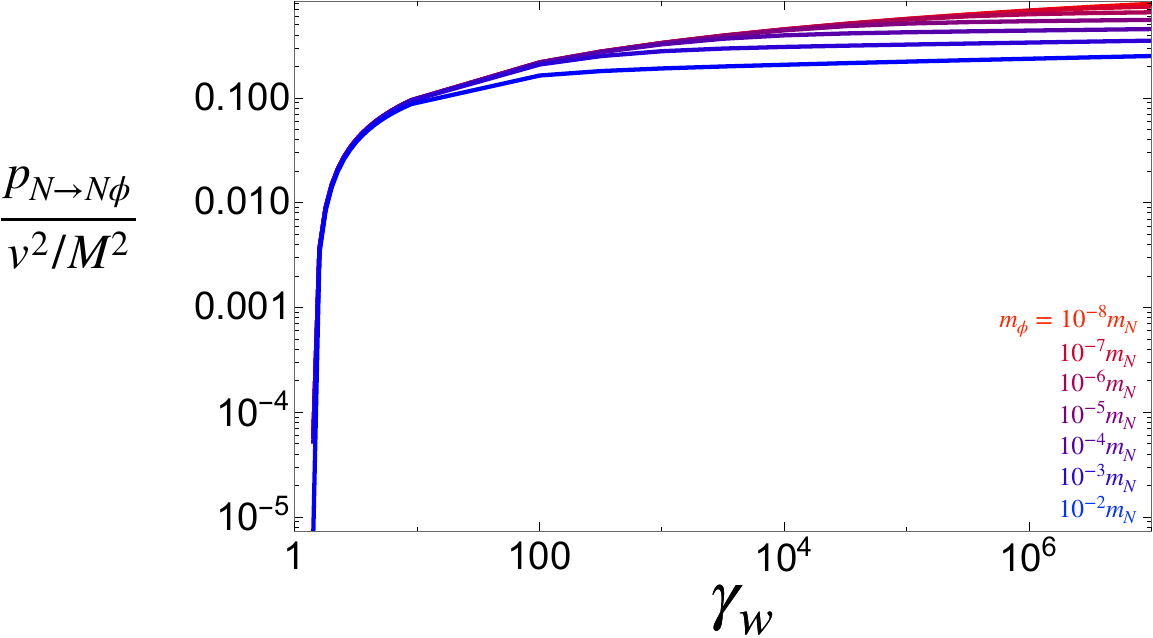}
  \end{center}
    \caption{The transition probability for $N\to N\f$ by varying the wall Lorentz factor. Here I choose $m_\phi=\{10^{-8}\cdots 10^{-2}\}m_N$ from top to bottom.  }
    \label{fig:1}
    \end{figure}

   \begin{figure}[t!]
  \begin{center}  
\includegraphics[width=140mm]{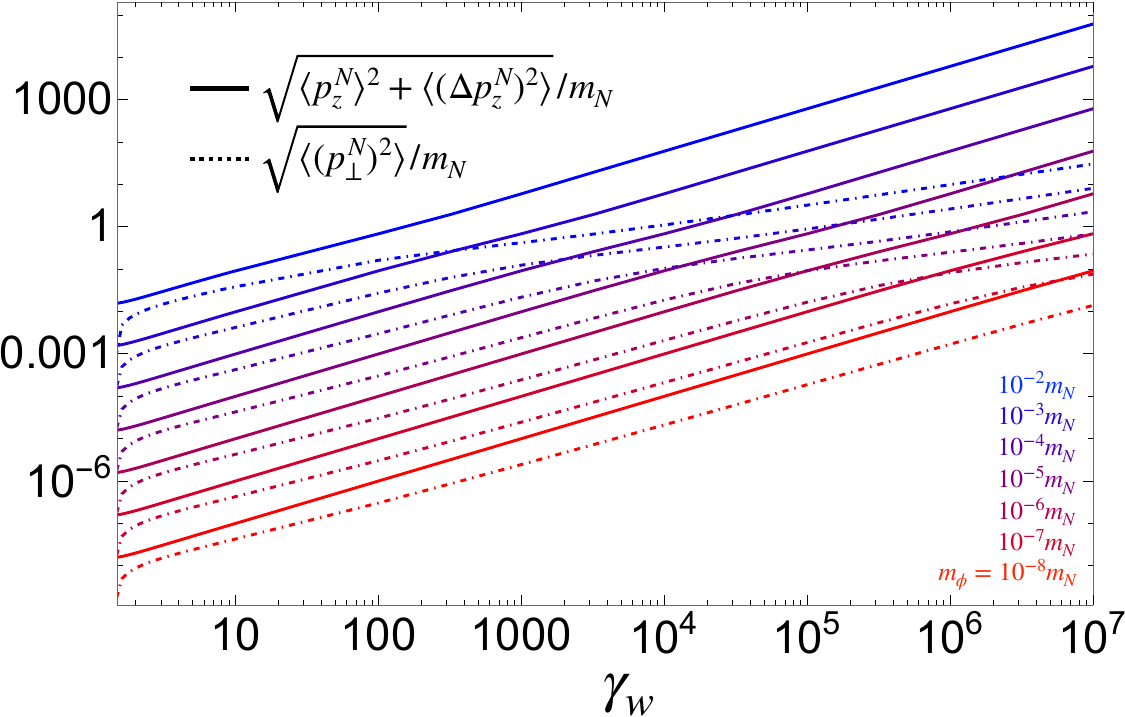}
\includegraphics[width=135mm]{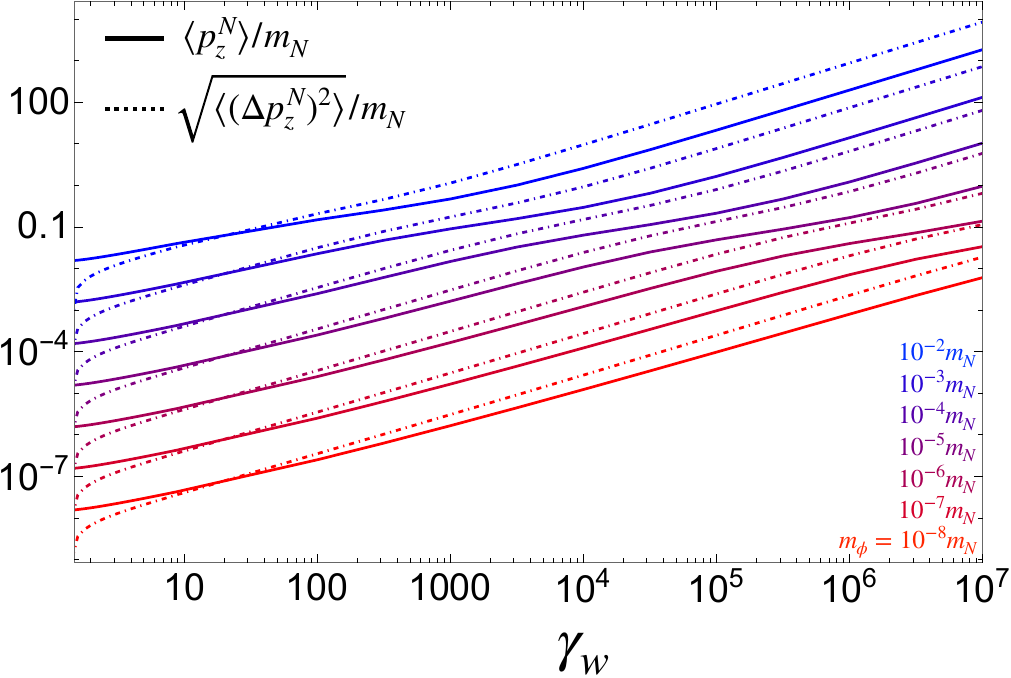}
  \end{center}
    \caption{The typical recoil momentum for $N\to N\f$ by varying the wall Lorentz factor. $m_\phi=\{10^{-8},\cdots 10^{-2}\}m_N$ from bottom to top. Solid (dashed) lines are for $\vev{p^{\rm Earth}_{f,z}}$ ($\sqrt{\vev{p_\perp^2}}$) in the top panel. In the bottom panels, solid (dashed) lines are $\vev{p^{\rm Earth}_{f,z}} (\sqrt{\vev{(\Delta p^{\rm Earth}_{f,z})^2}}$).  }
    \label{fig:2}
    \end{figure}

 $F\simeq\sqrt2\,v/M$ in the $\ab{\Delta p_z}\ll m_{\phi}$ limit of \Eq{F}.  
Substituting this constant and integrating over $x$ and $p_\perp^{2}$ in the region $\ab{\Delta p_z}<m_{\phi}$ give the production probability shown in Fig.\,\ref{fig:1}.  
The curves correspond to $m_{\phi}=\(10^{-8}\cdots10^{-2}\)\,m_N$ (top to bottom) and demonstrate that $P_{N\to N\phi}$ is $(0.1-1)\(\frac{v}{M}\)^2$ in most of parameter region, while it is  suppressed with $v_w\ll1$. 

Fig.~\ref{fig:2} (upper panel) displays the typical recoil momenta parallel $(\sqrt{\vev{p_{z,f}^{\rm Earth}}^2+\vev{(\Delta p_{z,f}^{\rm Earth})^2}})$ and transverse ($\sqrt{\vev{p^2_\perp}}$) to the wall; the lower panel shows the mean and variance of the $z$-component.  
The dependence on $m_{\phi}\g_w \AND \g_w$ indicates that the recoil spectrum retains direct information about the wall width and velocity.
In the range that $\g m_\f\gtrsim m_N$ we have the dominant typical momentum in the $z$-direction. This region however is only justified for the elementary particle. For the nuclei/nucleons, we need to have $\g m_\f m_N \lesssim \GEV^2$, where the former is the ``center-of-mass-energy" between $N$ and $\phi$ in the wall, and the latter is the QCD scale.

Although we may have some relevant constraint, e.g. from fifth-force, on the Yukawa interaction $f'(v)$, this can be easily evaded, see Ref.\,\cite{Azatov:2024auq} for a concrete symmetry-restoring model, where $f'(v)-f'(0)\neq 0, f'(v)=0$, and the associated bubble velocities.  
Also multi-field models with almost degenerate vacua can achieve an enhanced bubble nucleation even for heavy $m_{\phi}$, e.g.,~\cite{Niedermann:2019olb}. Given those discussions, I will not compare the constraint with other ones that can be evaded in this paper.

\subsection{Scaling domain walls}

A second class of cosmic walls consists of \emph{domain walls}.  
They appear after a $Z_{2}$ phase transition---first-order or second-order---in which different Hubble patches settle into degenerate vacua separated by a potential barrier~\cite{Kibble:1976sj,Vilenkin:1984ib,Vilenkin:2000jqa}; see also~\cite{Yin:2024pri} for variants without explicit symmetry restoration.  
Analogous walls also arise for axion fields without a $Z_{2}$ symmetry when large inflationary fluctuations drive neighbouring patches to field values that differ by $\O(v)$, the periodicity scale~\cite{Gonzalez:2022mcx,Kitajima:2023kzu}.  
In both cases, once the mass scale of the field becomes comparable to the expansion rate, the field begins to roll and stable walls form.

After formation, the network approaches a scaling solution with $\O(1)$ wall per Hubble volume at all times. Assuming the network formation much before the formation of the paleo detector, a wall crosses the detector only $\O(1)$ time in $13.8\,$Gyr.  
Typical wall velocities are semi-relativistic, with Lorentz factor $\gamma_{w}\sim1$.

\bef[t]
\bec
  \includegraphics[width=1\linewidth]{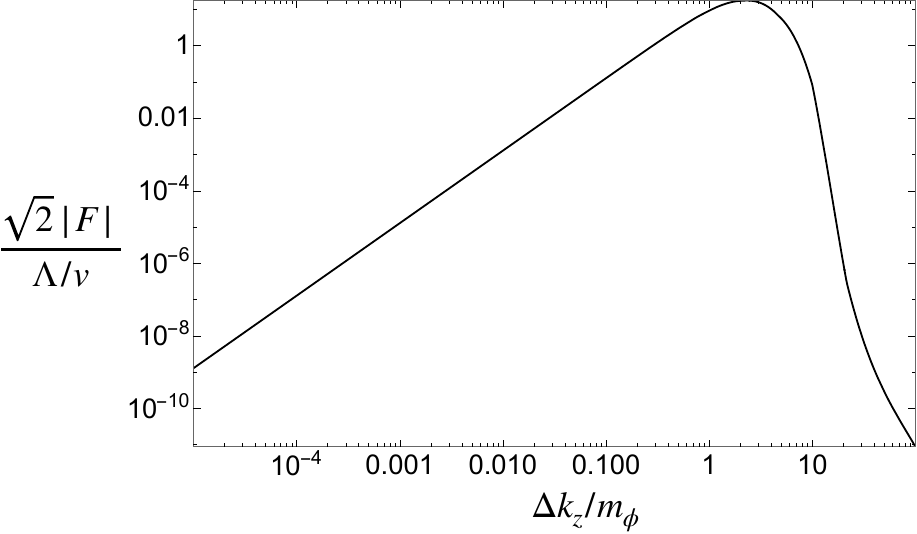}
\caption{$|F|$ for the axion profile $\f=(\tanh(m_{\phi}z)-1)/2$.}
\label{fig:3}
\eec
\eef

\bef[t!]
\bec
  \includegraphics[width=0.8\linewidth]{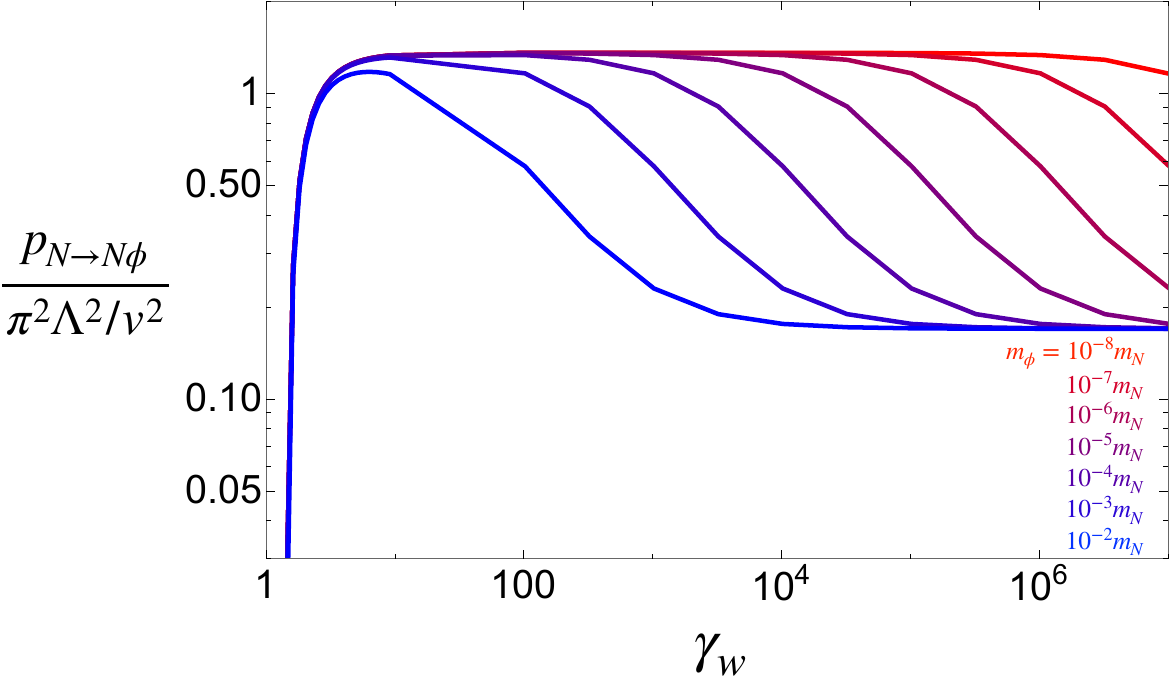}
\caption{As Fig.~\ref{fig:1} but for an axion wall with $\theta=0$.}
\label{fig:axion1}
\eec
\eef

\bef[t!]
\bec
  \includegraphics[width=0.8\linewidth]{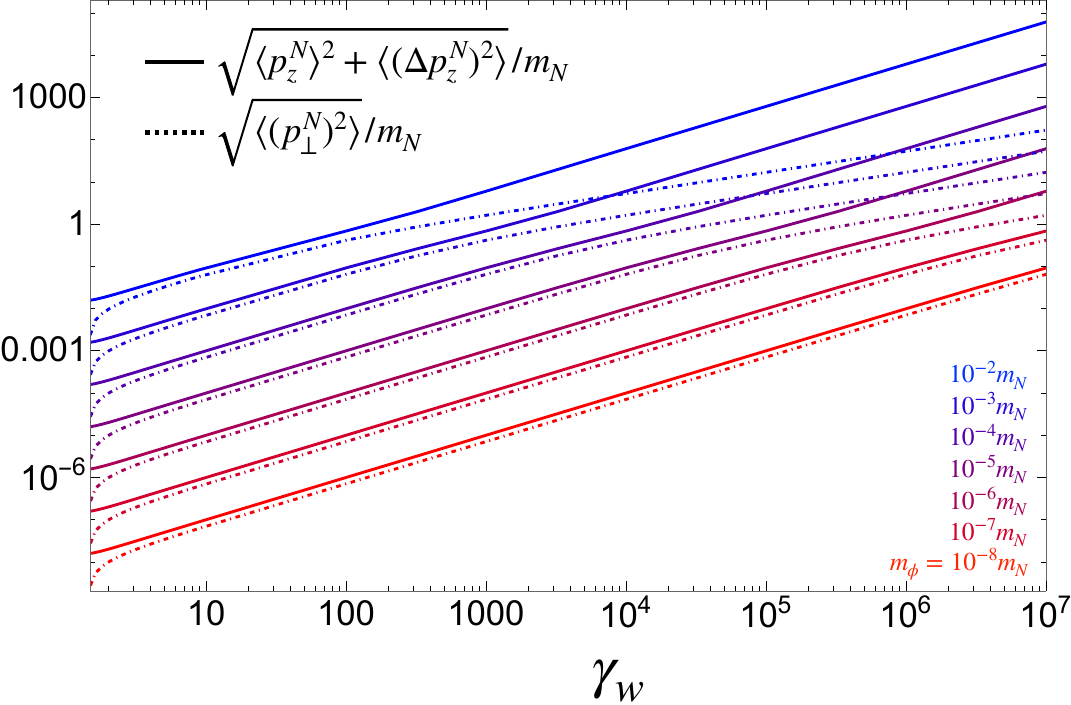}\\
  \includegraphics[width=0.77\linewidth]{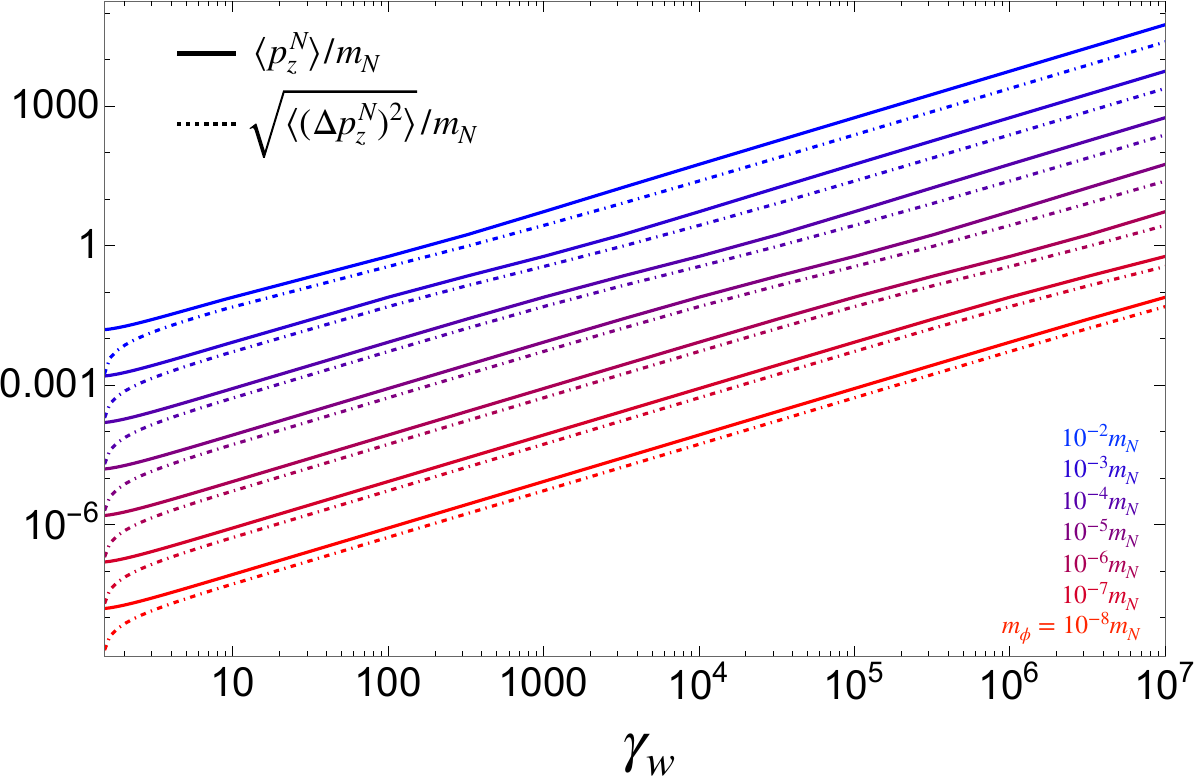}
\caption{As Fig.~\ref{fig:2} but for the axion wall with $\theta=0$.}
\label{fig:axion2}
\eec
\eef

Because the two vacua separated by the wall are equivalent, 
for an axion wall, which is my focus for concreteness,  the wall-fermion coupling must share that periodicity.  
We take
\beq
  f(\f)=\Lambda\cos\(2\pi \f/v + \theta \) ,
\eeq
so the interaction respects $\f\to\f+2\pi v$.  
The special choice $\theta=0$ preserves CP and automatically evades various limits such as  fifth-force.\footnote{%
  In a $Z_{2}$ model one could instead use $f(\f)=\f^{2}/M$ with boundary values $\f(\infty)=v$ and $\f(-\infty)=-v$.}
The filtering factor $F$ vanishes both for $\ab{\Delta p_{z}}\ll m_{\phi}$ and $\ab{\Delta p_{z}}\gg m_{\phi}$ in \Eq{F}.  
The magnitude $|F|$ for the benchmark profile is plotted in Fig.~\ref{fig:3}.

I stress that the qualitative discussion is unchanged from the FOPT case once the replacement
$,v/M \to 2\pi \Lambda/v,$ is made. The reason is twofold: (i) the filtering factor is non-zero for
$\Delta p_{z}\sim m_{\phi}$, and (ii) the phase-space integral over
$x$ and $p_{\perp}^{2}$ with $|\Delta p_{z}|\lesssim m_{\phi}$ gives a contribution of the same order as the region $\Delta p_{z}\sim m_{\phi}$.\footnote{For other reactions, e.g.\ soft‐photon emission, infrared phase space may dominate. Then, (ii) is not satisfied. }

Numerical results obtained by integrating over $|\Delta k_{z}|<m_{\phi}$ and using the $|F|$ profile of Fig.\,\ref{fig:3} are shown in Figs.\,\ref{fig:axion1} and~\ref{fig:axion2}. The $\g_{w}\gg 1$ behavour differs slightly from the FOPT curves, but the overall magnitudes remain comparable.

The difference originates from the more restricted phase space at
$\Delta p_{z}\sim m_{\phi}$: in the bubble-wall case the additional region
$\Delta p_{z}\ll m_{\phi}$ is allowed, enlarging the phase space for
$p_{f}^{z}$. Consequently the variance
$\vev{\bigl(\Delta p_{f,z}^{\text{Earth}}\bigr)^{2}}$ is modestly larger for the FOPT wall than for the scaling domain wall.

\subsection{Prospects and limits for cosmic walls with paleo detectors}

A rough rule of thumb is
\beq
  P_{N\to N\phi}\;\sim\;0.1\ab{f'_{\text{typ}}}^{2},
\eeq
with $f'_{\text{typ}}=\sqrt2\,v/M$ for late FOPT walls and $f'_{\text{typ}}=\sqrt2\pi\L/v$ for scaling walls.  This is good approximation when the $\gamma_w$ is larger than $\O(1)$ but not too large as seen in Figs. \ref{fig:1} and \ref{fig:axion1}. 
The typical injected momentum is
\beq
  p_{N}^{\rm Earth}\;\sim\;\g_w\,m_{\phi},
\eeq
i.e.\ the inverse wall width in the Earth frame.

For a target containing $N_{\text{target}}$ scatterers (e.g. 100g muscovite), the expected number of events is
\beq
  N_{\text{event}}\;=\;P_{N\to N\phi}\,N_{\text{target}}
  \;\sim\;
  10\;
  \left(\frac{\ab{f'_{\text{typ}}}}{10^{-11}}\right)^{2}
  \frac{N_{\text{target}}}{10^{24}} .
\eeq

Current and future paleo read-out thresholds correspond to recoil energies $0.1\KEV$--$100\KEV$~\cite{Edwards:2018hcf,Baum:2018tfw,Drukier:2018pdy}), i.e.\ momentum sensitivity
\beq
  p_{N}^{\text{sens}}\simeq(5\MEV\text{--}50\MEV)\,
  \sqrt{\frac{m_{N}}{30\GEV}} .
\eeq
Hence the detector is sensitive when
\beq
  m_{\phi}\,\g_w\;\sim\;
  (5\MEV\text{--}50\MEV)\,
  \sqrt{\frac{m_{N}}{30\GEV}},
\eeq
achievable for $m_{\phi}\sim\MEV$ with $\g_w\sim1$ or for smaller $m_{\phi}$ with larger $\g_w$.

Using the WIMP-search data obtained with ancient muscovite mica of $0.5$Gyr old in Ref.\,\cite{Snowden-Ifft:1995zgn}, I recast the event count into a bound on cosmic walls passing in the last 0.5 Gyr.  
The resulting $90\%$ CL limit on the interaction is shown in Fig.\,\ref{fig:limit}; the four lines correspond to
$N=$$^{16}\mathrm{O}$,
$^{27}\mathrm{Al}$,
$^{28}\mathrm{Si}$,
and
$^{39}\mathrm{K}$,
respectively, from left to right. 
To translate the recoil spectrum I adopt the approximated formulas of this subsection and assume a dark matter velocity
$v=220\;\mathrm{km\,s^{-1}}$.

This bound can be tightened—possibly even turned into a detection—by a dedicated re-analysis with (i) the full model-dependent recoil distribution and (ii) newer high-resolution track data, should it become available.  
Track geometry itself is also informative.  
For non-relativistic walls ($v_{w}\ll1$) and the relativistic $N$ the transverse kick dominates (upper panels of Figs.\,\ref{fig:2} and \ref{fig:axion2}); event tracks all across the Earth would then appear nearly parallel,\footnote{The wall transit time is so short for $m_\f\gg 10^{-15}\EV$ that Earth’s rotation and revolution can be neglected.} providing a smoking-gun signature beyond the simple rate estimate.  
For $\g_{w}\gtrsim1$ and non-relativistic $N$ the distribution broadens, but precise mapping of track orientations can still reveal the wall’s velocity and direction (Figs.\,\ref{fig:2} and \ref{fig:axion2}).

\bef[t]
\bec
\includegraphics[width=1.00\linewidth]{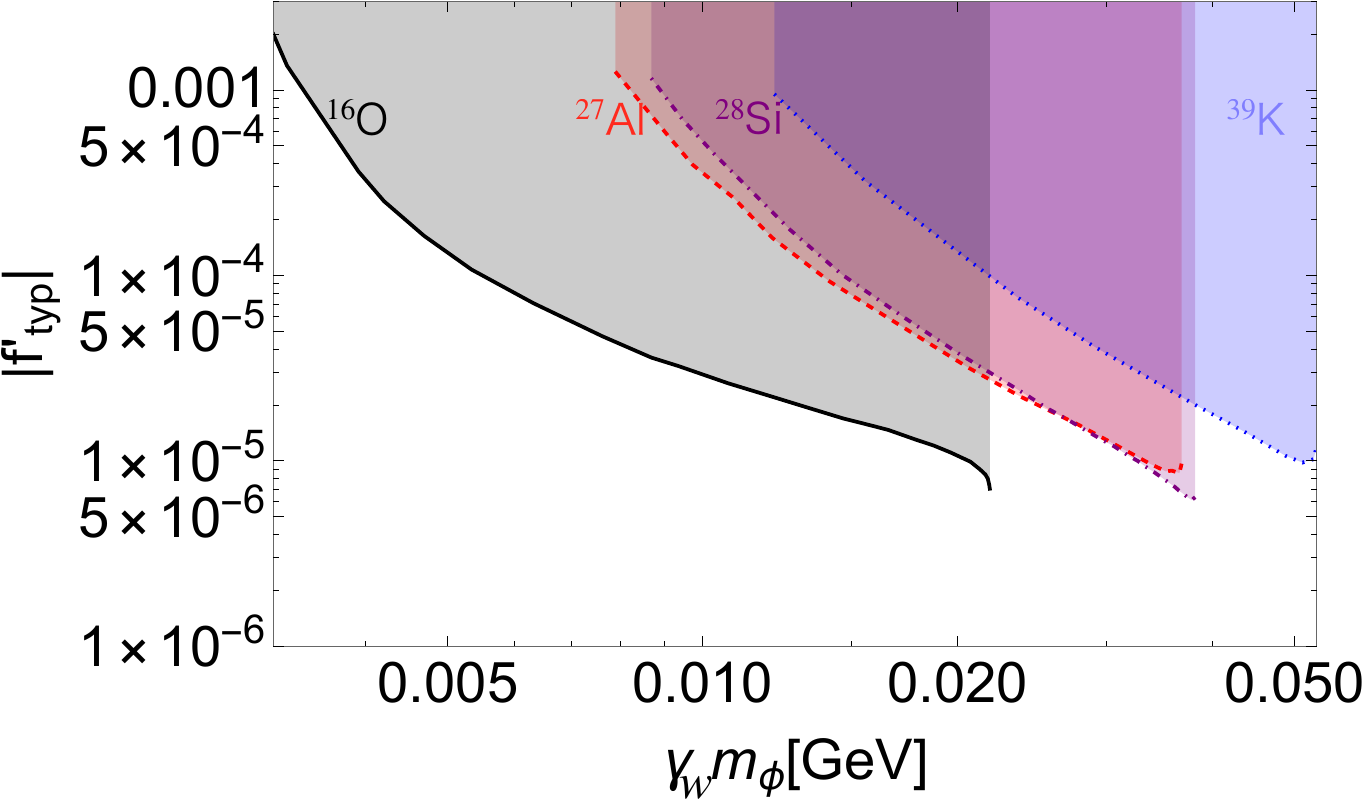}
\caption{Direct detection limit on cosmic walls derived from the muscovite data of Ref.\,\cite{Snowden-Ifft:1995zgn}. The limit applies to a wall passage occurring within the last 0.5\,Gyr. }
\label{fig:limit}
\eec
\eef

\section{Conclusions and discussion}

Cosmic walls intersect Earth at most $\O(1)$ time over the age of the Universe, and each transit lasts far less than a second.  
If the wall--matter coupling generates recoil energies above the paleo detector threshold, every mineral that already existed at the time of passage acquires damage tracks.

This gives a clear signature: minerals older than the wall event should contain  tracks, whereas younger samples should be free of them.  
The three-dimensional orientation of the tracks further encodes the wall’s Lorentz factor and trajectory.

In this work I showed that paleo detectors enable a truly direct search for cosmic walls and derived an initial limit by recasting the existing muscovite data.  
A dedicated re-analysis with modern, high-resolution read-out—or new data—could improve the limit or discover a signal.

Ultra-relativistic walls also provide an indirect handle.  
Their large momentum transfer can boost Standard-Model or hidden particles throughout cosmic space, effectively making the walls a source of cosmic rays.  
For a typical injected momentum $p_{X}\sim m_{\phi}\g_{w}$ the flux is estimated as
\beq
  {\rm flux}\;\simeq\;
\,P_{N\to X\phi}\,n
  \;\simeq\;
  10\ {\rm km}^{-2}\,{\rm yr}^{-1}\,
  \left(\frac{P_{N\to X\phi}}{10^{-20}}\right)
  \left(\frac{n}{10^{-7}\,{\rm cm}^{-3}}\right),
\eeq
where $n$ is the average baryon number density.  Here $X$ denotes a correct of the particles since the reaction is far beyond the QCD or even weak scale. $N$ may be also neutrinos.
Such a flux could be sought in large-area or large-volume observatories such as Telescope Array~\cite{TelescopeArray:2018xyi}, Pierre Auger~\cite{PierreAuger:2020kuy}, IceCube~\cite{IceCube:2018fhm}, or KM3NeT~\cite{KM3NeT:2025npi}, depending on the particle species $X$.

Taken together, track geometry in ancient minerals and possible cosmic-ray signatures offer complementary avenues to probe or constrain late-time cosmic walls.

\section*{Acknowledgement}
I am grateful to the organizers of the Mineral Detection of Neutrinos and Dark Matter 2025 workshop for inviting me to give a talk, which inspired this study.  
This work is supported by JSPS KAKENHI Grant Nos. 22K14029, 22H01215, and by Selective Research Fund for Young Researchers from Tokyo Metropolitan University.

\appendix 
\section{Estimation for bosonic nuclei}

I now consider the possibility that a nucleus itself behaves as a boson.  
For simplicity I treat it as a complex scalar field and denote it by $N$.  
Its free Lagrangian contains the usual mass term
\beq
  \mathcal{L} \supset -\,m_{N}^{2}\,|N|^{2},
\eeq
where $m_{N}\simeq A\,m_{n}$ with $A$ the mass number and $m_{n}$ the nucleon mass.

Assume that the wall field couples universally to individual nucleons through
$f^{(n)}(\f)\,\bar n n$.  
Then, to leading order in $A$, the induced coupling to the composite scalar $N$ is
\beq
  \mathcal{L}_{\mathrm{int}}
  \supset 2\,m_{N}\,A\,f^{(n)}(\f)\,|N|^{2}
  \simeq 2\,m_{N}\,f(\f)\,|N|^{2},
\eeq
where I have defined $f(\f)=A\,f^{(n)}(\f)$ in analogy with the fermionic case.  
The squared matrix element becomes
\beq
  |\,\mathcal{M}\,|^{2}\;\simeq\;(2m_{N})^{2}.
\eeq

\begin{figure}[t]
  \centering
  \includegraphics[width=140mm]{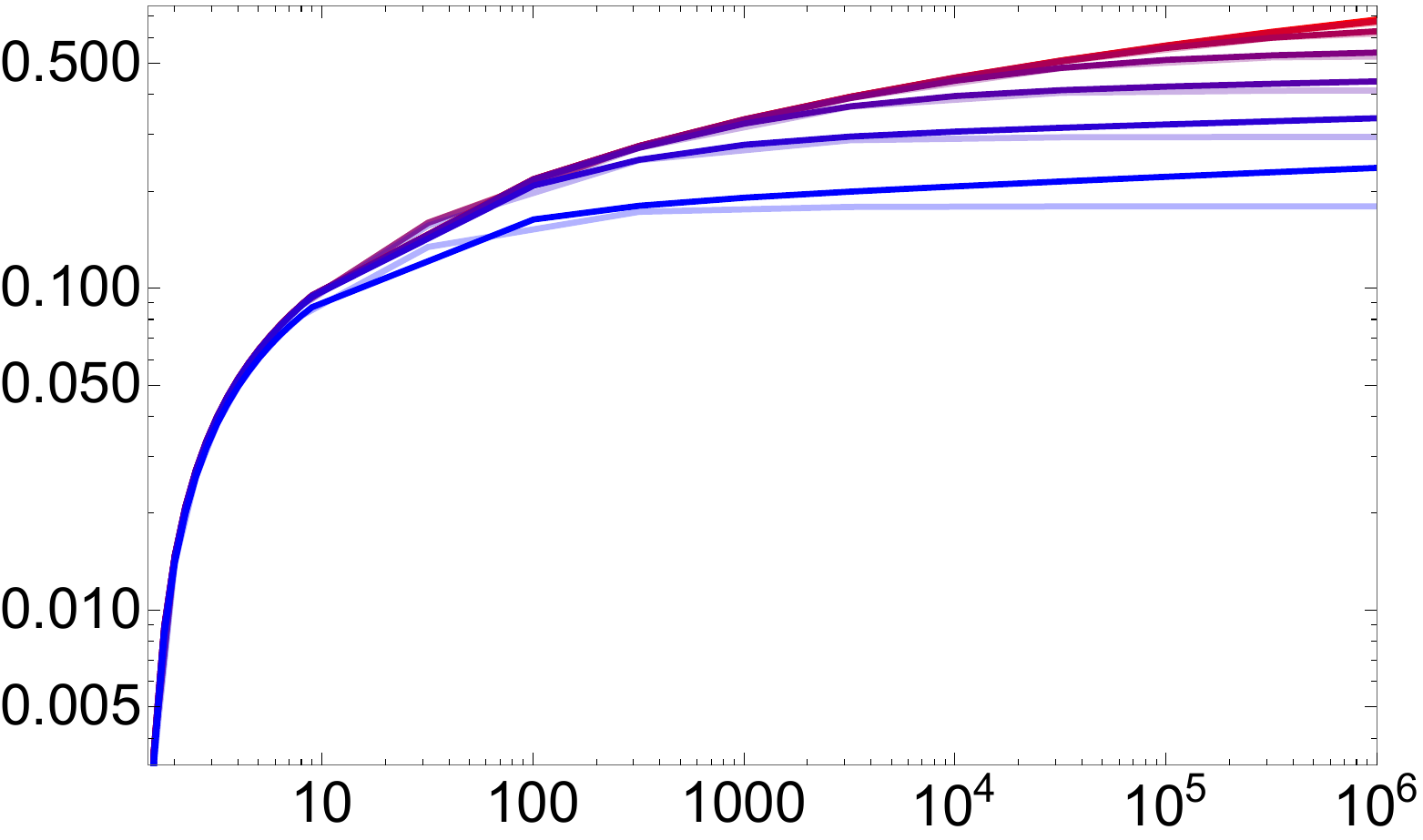}
  \caption{Production probability for bosonic nuclei (darker curves) compared with the fermionic case of Fig.~\ref{fig:1} (lighter curves).}
  \label{fig:4}
\end{figure}

\begin{figure}[t]
  \centering
  \includegraphics[width=140mm]{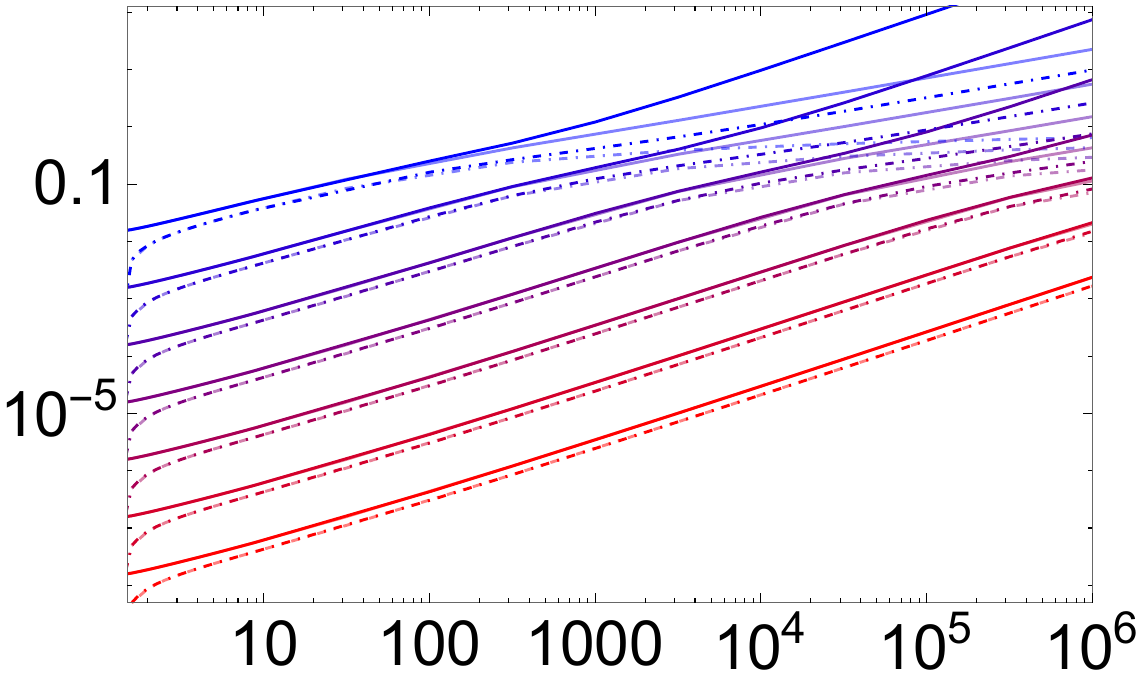}\\
  \includegraphics[width=135mm]{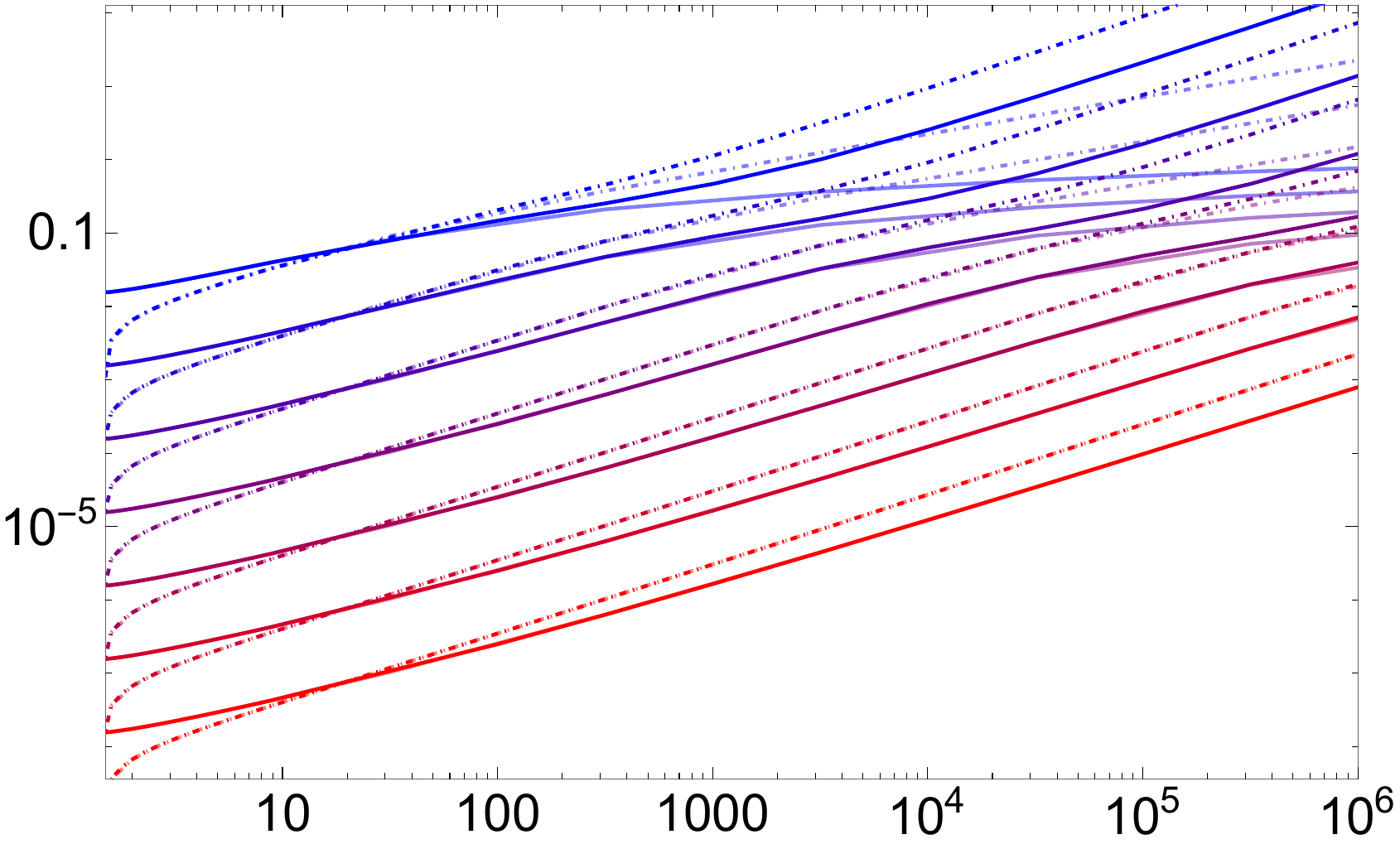}
  \caption{Typical recoil momenta for bosonic nuclei (darker curves) versus the fermionic results of Fig.~\ref{fig:2} (lighter curves).}
  \label{fig:5}
\end{figure}

Repeating the analysis of the FOPT wall with this replacement yields the results shown in Figs.~\ref{fig:4} and~\ref{fig:5}.  
For comparison the fermionic curves from Figs.~\ref{fig:1} and~\ref{fig:2} are overlaid (lighter color).  
The only significant deviation appears when the typical momentum approaches $m_{N}$, where the effective field treatment breaks down and a microscopic description would be required.

\bibliography{Paleo2.bib}

\end{document}